\documentclass[prd,a4paper,twocolumn,preprintnumbers,floatfix,showkeys,showpacs,aps,nofootinbib]{revtex4}
\usepackage{latexsym}
\usepackage{epsfig}

\newcommand{\beq}{\begin{eqnarray}}
\newcommand{\eeq}{\end{eqnarray}}
\newcommand{\be}{\begin{eqnarray}}
\newcommand{\ee}{\end{eqnarray}}

\begin{document}

\title{Dark Energy Anisotropic Stress and Large Scale Structure Formation}

\date{\today}

\author{Tomi Koivisto$^1$\footnote{tomikoiv@pcu.helsinki.fi} and David F. Mota$^2$\footnote{mota@astro.uio.no}}
\affiliation{$^1$ Helsinki Institute of Physics,FIN-00014 Helsinki, Finland}
\affiliation{$^2$ Institute of Theoretical Astrophysics, University of Oslo, Box 1029, 0315 Oslo, Norway}

\begin{abstract}

We investigate the consequences of an imperfect dark energy fluid on the large 
scale structure. A phenomenological three parameter fluid description is 
used to study the effect of dark energy on the cosmic microwave background 
radiation (CMBR) and matter power spectrum. 
In addition to the equation of state and the sound speed, we
allow a nonzero viscosity parameter for the fluid. 
Then anisotropic stress perturbations are generated in dark energy. 
In general, we find that this possibility is not excluded by the present
day cosmological observations. In the 
simplest case when all of the three parameters are constant, we find that 
the observable effects of the anisotropic stress can be closely
mimicked by varying the sound speed of perfect dark energy. However, now
also negative values for the sound speed, as expected for adiabatic fluid
model, are tolerable and in fact could explain the observed low quadrupole
in the CMBR spectrum. We investigate also structure formation of 
imperfect fluid dark energy characterized by an evolving equation of state. 
In particular, we study models unifying dark energy with dark matter,
such as the Chaplygin gas or the Cardassian expansion, 
with a shear perturbation included. This can stabilize the growth
of inhomogeneities in these models, thus somewhat improving their 
compatibility with large scale structure observations. 

\end{abstract}

\keywords{Cosmology: Theory, Large-Scale Structure of Universe}
\pacs{98.80.-k,98.80.Jk}

\maketitle

\section{Introduction}

Dark energy is a fundamental component of the nowadays standard 
cosmological model.
It would be  very difficult to explain the set of present days 
cosmological observations without it. 
Specifically, we refer to the luminosity-redshift 
relationship from observations of supernovae of type Ia (SNIa)   \cite{Riess:1998cb,Perlmutter:1998np,Riess:2004nr}, 
the matter power spectrum of large scale structure as inferred 
from galaxy redshift surveys like the Sloan Digital Sky Survey (SDSS)   \cite{Tegmark:2003ud} 
and the 2dF Galaxy Redshift Survey (2dFGRS)   \cite{Colless:1998yu}, 
and the anisotropies  in the Cosmic Microwave Background Radiation (CMBR)   \cite{Spergel:2003cb}.

Despite of its major importance in explaining the astrophysical data, 
the nature of dark energy is one of the greatest mysteries of modern cosmology. 
The simplest and most popular candidates for it are the cosmological 
constant (see e.g.   \cite{Carroll:2000fy}), and minimally coupled 
scalar fields (see e.g.  
 \cite{Wetterich:1994bg,Ratra:1987rm,Caldwell:1997ii,Zlatev:1998tr}).
However many other candidates were proposed based on high energy physics phenomenology  ( see e.g. 
\cite{Amendola:1999er,Farrar:2003uw,Brookfield:2005td,Bagla:2002yn,Padmanabhan:2002cp,Armendariz-Picon:2000ah,Bertolami:1998dn,Boisseau:2000pr,
Caldwell:1999ew,Gibbons:2003gb,Chiba:1999ka}),
and many investigations on  their possible astrophysical and cosmological signature were undertaken
( see e.g.   
\cite{Seljak:2004xh,Abramo:2004ji,Mota:2003tc,Evans:2004iq,Manera:2005ct,Amarzguioui:2004kc,
Alam:2003fg,Mota:2003tm,Melchiorri:2002ux,Mota:2004pa,Hannestad:2002ur,
Koivisto:2004ne,Nunes:2004wn,Koivisto:2005nr} ).

With so many possible candidates it is 
imperative to understand what are the main properties of the dark 
energy component that could have specific signatures in 
the astronomical data, and so could help us to discriminate among 
all these models.

In a phenomenological approach, dark energy might be mainly 
characterized by its equation of state $w$, its sound speed $c_s$, and its anisotropic stress $\sigma$   \cite{Hu:1998tj}.
Much effort has been put into determining the equation of state of
dark energy, in an attempt to constrain theories. The 
equation of state determines the decay rate of energy and thus
affects both the background expansion and the evolution of
matter perturbations (see e.g   \cite{Peebles:2002gy}).
An equally insightful characteristic of dark
energy is its speed of sound. This does not affect the
background evolution but is fundamental in characterizing the behavior 
of its perturbations. Hence many authors have explored its effect on
the evolution of fluctuations in the matter distribution 
( see e.g.   \cite{Bean:2003fb,Sandvik:2002jz,Avelino:2002fj,Balakin:2003tk}).
However, the investigation of the effects of the 
anisotropic stress has been widely neglected. 

The main reason for disregarding the anisotropic stress in the dark energy 
fluid might be that conventional dark energy candidates, such as the 
cosmological constant or scalar fields,  
are perfect fluids with $\sigma=0$.  However, since there is no fundamental theoretical model to 
describe dark energy, there are no strong reasons to stick to such assumption. 
In fact, dark energy vector field candidates have been proposed
  \cite{Armendariz-Picon:2004pm,Kiselev:2004py,Zimdahl:2000zm,
Novello:2003kh,Wei:2006tn}, and these have $\sigma\neq 0$.
Of course, if dark energy  is such a vector, one might  break   the   isotropy   of   a
Friedmann-Robertson-Walker universe.  However, as long as it
remains subdominant,  this violation  is likely to  be observationally
irrelevant    \cite{Barrow:1997as}.   Once  dark  energy  comes  to  dominate
though, one would expect an  anisotropic expansion of the universe, in
conflict  with  the  significant  isotropy  of  the  CMBR   \cite{Bunn:1996ut}. But on the other hand 
there appears to be hints of statistical anisotropy in the CMBR 
fluctuations  \cite{Jaffe:2005pw,Bielewicz:2004en,Larson:2004vm,Schwarz:2004gk,Copi:2003kt,deOliveira-Costa:2003pu}.  

Recently the possibility of viscous dark energy has 
gained attention  \cite{Brevik:2004sd,Brevik:2005ue,Nojiri:2005sr,
Brevik:2005bj}. These models are usually restricted to the context of bulk
viscosity, although one could expect the shear viscosity to be 
dominant  \cite{Brevik:2005bj}. One can allow bulk viscosity in a 
Friedmann-Robertson-Walker (FRW) universe, but when the shear is not 
neglected one has to face the difficulties of an anisotropic universe. 
However, shear viscosity at the perturbative level is compatible with the 
assumption of an isotropic FRW background. In fact the anisotropic stress 
perturbation is crucial to the understanding of evolution of 
inhomogeneities in the early, radiation dominated universe.  
Therefore an obviously interesting question is whether present 
observational data could allow for an anisotropic stress perturbation in the 
late universe which is dominated by the mysterious dark energy fluid. 
 
Motivated by all these possibilities, we investigate if 
the possible existence of an anisotropic stress in the dark 
energy component would result in a specific 
cosmological signature which could be probed using large scale structure 
data, and if it would still be compatible with the latest CMBR temperature 
anisotropies and the matter power spectrum.

The article is organized as follows: In section II we discuss the 
parameters describing a general dark energy fluid with anisotropic 
stress. In section III we consider dark energy imperfect fluid 
models parameterized with a constant equation of state, sound speed and anisotropic stress. 
We investigate the effects on the late time perturbation 
evolution, in the integrated Sachs-Wolfe (ISW) effect of 
the CMBR anisotropies and on the matter power spectrum. In section IV we 
extend the analysis to models unifying dark energy with dark matter. We 
end the article with a summary of our findings and 
conclusions. 

\section{Dark Energy Stress Parameterization}

In its simplest descriptions the dark energy component is described
fully by its equation of state, defined as
\be
w \equiv \frac{p}{\rho},
\ee
where $\rho$ is the energy density and $p$ is the pressure of the fluid.
If $u_\mu$ is the four-velocity of the fluid, and the projection
tensor $h_{\mu\nu}$ is defined as $h_{\mu\nu} \equiv 
g_{\mu\nu} + u_\mu u_\nu$, we can write the energy momentum tensor for a 
general 
cosmological fluid as
\be \label{fluid}
T_{\mu\nu}= \rho u_\mu u_\nu + ph_{\mu\nu} + \Sigma_{\mu\nu},
\ee
where $\Sigma_{\mu\nu}$ can include only spatial inhomogeneity. We define perfect fluid by 
the condition $\Sigma_{\mu\nu}=0$.
If in addition the fluid is adiabatic, $p=p(\rho)$, the evolution of 
its perturbations is described by the adiabatic speed of sound $c_a$. 
This is in turn fully determined by the equation 
of state $w$, 
\be
\label{ca}
c_a^2 \equiv \frac{\dot{p}}{\dot{\rho}} = w - \frac{\dot{w}}{3H(1+w)}.
\ee
For an adiabatic fluid, $\delta p = c_a^2 \delta \rho$.

In the general case, there may be more degrees of freedom and 
the pressure $p$ might not be a unique function of the energy density 
$\rho$. An extensively studied example is quintessence 
 \cite{Wetterich:1994bg,Ratra:1987rm,Caldwell:1997ii,Zlatev:1998tr}.
For such a scalar field the variables $w$ and $c_s^2$ depend on two 
degrees of freedom: the field and its derivative, or equivalently, the kinetic and 
the potential energy of 
the field. Then the dark energy (entropic) sound speed is defined as the 
ratio of pressure and 
density perturbations in the frame comoving with the dark energy fluid,
\be\label{cs}
c_s^2 \equiv \frac{\delta p}{\delta \rho}_{|de}.
\ee
In the adiabatic case, $c_s^2= c_a^2$, which holds in any frame, but
in general the ratio $\delta p/\delta \rho$ is gauge dependent.  Hence, 
in the case of entropic fluid such as scalar fields, one 
needs both its equation of state and its sound speed as defined in 
Eq. (\ref{cs}), to have a complete description of dark energy 
and its perturbations.

However, in order to have an even more general set of parameters to fully 
describe a dark energy fluid and its perturbations, besides $w$ and 
$c_s$, one should also consider the possibility of anisotropic stress. 
This is important because it enters directly into the Newtonian metric, 
as opposed to $w$ and $c_s$ which only contribute through the causal 
motion of matter  \cite{Hu:1998tj}. 

Taking this generalization into account, in the synchronous 
gauge  \cite{Ma:1995ey}, the evolution equations for the 
dark energy density perturbation and velocity potential can be written 
as  \cite{Hannestad:2005ak}
\beq \label{deltaevol}
\dot{\delta} &=& 
-(1+w)\left\{\left[k^2+9H^2(c_s^2-c_a^2)\right]\frac{\theta}{k^2}
               + \frac{\dot{h}}{2}\right\} \nonumber\\ &-& 3H(c_s^2-w)\delta,
\eeq
\be \label{thetaevol}
\dot{\theta} = -H(1-3c_s^2)\theta+\frac{c_s^2k^2}{1+w}\delta-k^2\sigma,
\ee
where $h$ is the trace of the synchronous metric perturbation.
Here $\sigma$ is the anisotropic stress of dark energy, related to notation
of Eq.(\ref{fluid}) by $(\rho + p)\sigma \equiv 
-(\hat{k}_i\hat{k}_j-\frac{1}{3}\delta_{ij})\Sigma^{ij}$. Basically, while 
$w$ and $c_s^2$ determine respectively the background and perturbative 
pressure of the fluid that is rotationally invariant, $\sigma$ quantifies 
how much the pressure of the fluid varies with direction.

Generally such a property implies shear viscosity in the fluid, and thus 
its effect is to damp perturbations. A covariant form for the 
viscosity generated in the fluid flow is \cite{misner}
\be \label{l-l}
\Sigma_{\mu\nu} = \varsigma\left(u_{\mu;\alpha}h^\alpha_\nu +u_{\nu;\alpha}h^\alpha_\nu 
- u^\alpha_{\phantom{\alpha};\alpha} h_{\mu\nu}\right) 
+ \zeta u^\alpha_{\phantom{\alpha};\alpha} h_{\mu\nu}.
\ee
Now the the conservation equations $T^{\mu \nu}_{\phantom{\mu\nu};\mu}=0$ 
reduce to the Navier-Stokes equations in the non-relativistic limit. Here 
$\varsigma$ is the shear viscosity coefficient, and $\zeta$ represents
bulk viscosity. Here we set the latter to zero since we demand that 
$\Sigma_{ij}$ is traceless. In cosmology we have $u_\mu = (1,-v_{,i})/a$ in the
synchronous gauge, and the velocity potential $\theta$ is the divergence of the 
fluid velocity $v$. It is then straightforward to check that the components
of Eq.(\ref{n-s}) vanish except in the off-diagonal of the perturbed spatial
metric. One finds that
\be \label{n-s}
\sigma = \frac{\varsigma}{k}\left(\theta - \dot{H}_T\right),  
\ee
where $H_T$ is the scalar potential of the tensorial metric perturbations, 
which in the synchronous gauge equals $H_T=-h/2-3\eta$, where $\eta$ is 
a metric perturbation. From the coordinate 
transformation properties of $T_{\mu\nu}$ it follows that $\sigma$ 
must be gauge-invariant, 
and indeed the linear combination $\theta-\dot{H}_T$ is frame-independent. 

However, the anisotropic stress is not necessarily given 
directly by $\theta-\dot{H}$. For neutrinos this term instead acts as a 
source for the anisotropic stress, which is also coupled to higher 
multipoles in the Boltzmann hierarchy. Thus the evolution of the stress 
must, at least in principle, be solved from a complicated system of 
evolving multipoles. 

The approach we will use in this article to specify 
the shear viscosity of the fluid is more in line with the neutrino stress 
than Eq.(\ref{n-s}). Following Hu  \cite{Hu:1998kj}, we describe the 
evolution of the anisotropic stress with the equation
\be \label{sigmaevol}
\dot{\sigma}+3H\frac{c_a^2}{w}\sigma =
\frac{8}{3}\frac{c_{vis}^2}{1+w}(\theta+\frac{\dot{h}}{2}+3\dot{\eta}).
\ee
Then the shear stress is not determined algebraically from fluctuations
in the fluid as was the case in Eq. (\ref{n-s}), but instead it
must be solved from a differential equation.

This phenomenological set-up is motivated as follows  \cite{Hu:1998tj}. 
One can guess that the anisotropic stress is sourced by shear 
in the velocity and in the metric fluctuations. 
Again one must take into account the coordinate 
transformation properties of $\sigma$, and construct a 
gauge-invariant source term in the differential equation. 
As mentioned, an appropriate linear combination is  
$\theta-\dot{H}_T$. 
Up to the viscosity parameter $c^2_{vis}$, this determines the 
right hand side of Eq. (\ref{sigmaevol}). In the left hand side there appears 
also a drag term accounting for dissipative effects. We have 
adopted a natural choice for the dissipation time-scale, $\tau^{-1}_\sigma 
= 3H$. 

One may then check that Eq. (\ref{sigmaevol}) with $w=c_{vis}^2=1/3$ reduces 
to the evolution equation for the massless neutrino quadrupole in the truncation scheme 
where the higher multipoles are neglected  \cite{Ma:1995ey} (this applies 
also to photons when one ignores their polarization and coupling to 
baryons). In what follows, we will study the consequences of 
Eq.(\ref{sigmaevol}) for fluids with negative equations of state. 
For $w<-1$, one should consider negative values of $c_{vis}^2$, 
as was suggested in Ref.  \cite{Huey:2004jz}. So the parameter
$c_{vis}^2/(1+w)$ should remain positive. We will return to this 
in the section III.C.    
  
Note that the parameterization of Eqs.(\ref{deltaevol}), (\ref{thetaevol}) 
and (\ref{sigmaevol}) describes cosmological fluids in a very general way. 
The system reduces to cold dark matter equations 
when $(w,c_s^2,c_{vis}^2)$ is $(0,0,0)$ and relativistic 
matter corresponds to $(1/3,1/3,1/3)$. A scalar field with a canonical 
kinetic term is given by $(w(a),1,0)$, where $-1 < w(a) < 1$. With an 
arbitrary kinetic term one can construct k-essence models 
\cite{Armendariz-Picon:2000ah} characterized
by unrestricted equations of state and speeds of sound, 
$(w(a),c_s^2(a),0)$, but vanishing shear. 
On the other hand, one should keep in mind that the parameterization
cannot be completely exhaustive. It does not cover, for example, 
a cosmological fluid with anisotropic stress determined by 
Eq. (\ref{n-s}) when $\varsigma \neq 0$. We might address the viability
of this approximation elsewhere, but restrict here to the parameterization
given in Eq.(\ref{sigmaevol}). 

\section{Stressed Dark Energy Fluid}

We will investigate the effect of dark energy perturbations on the CMBR 
anisotropies and on the matter power spectrum with the simplest assumption 
that all of the three parameters $w$, $c_s^2$ and $c_{vis}^2$ are 
constant. 
This is an accurate description for a wide variety of models for which
these parameters can be well approximated at moderate redshifts by their 
time-averaged values. 
  
For the energy content of the universe we use $\Omega_b=0.044$, $\Omega_{cdm}=0.236$,
$\Omega_{de}=0.72$ and $h_0 = 0.68$. In all the numerical calculations we 
assume a scale invariant initial power spectra and set 
optical depth to last scattering to zero. We normalize the perturbation 
variables by setting the primordial comoving curvature perturbation to 
unity. 
To compare with observations, the resulting power spectra must then be 
multiplied by the primordial amplitude of the curvature perturbation. For 
this we employ the same normalization for all the models considered in 
this article. Although we do not search for the best-fit 
models here, we include the WMAP 
data  \cite{Spergel:2003cb} and the SDSS data  \cite{Tegmark:2003ud} in to 
the 
figures in order to give an idea about the viability of the studied models. 
The WMAP error bars include the cosmic variance, which is the 
dominant source of uncertainty for the small $\ell$'s. The calculations are 
performed with a modified version of the CAMB code   \cite{Lewis:1999bs}.

In addition to specifying the cosmological parameters and the evolution 
equations (\ref{deltaevol}), (\ref{thetaevol}) and (\ref{sigmaevol}),
we must address the initial conditions in the early universe. For
the relative entropy between radiation and dark energy to vanish, one must 
impose
\be \label{adi}
\delta & = & \frac{3}{4}(1+w)\delta_r, 
\nonumber \\ 
\theta & = & 
\frac{1}{1+9\frac{H^2}{k^2}(c_s^2-c_a^2)}\left[\theta_r-\frac{9}{4}H(c_s^2-c_a^2)\delta_r\right].
\ee
When there is no inherent entropy in the dark energy fluid, i.e. $c_s^2=c_a^2$,
adiabaticity means that the velocity potentials of all cosmic fluids are equal.
On the other hand, when $c_s^2 \neq c_a^2$, we see that the above condition
implies that the dark energy velocity potential is negligible in the early
universe, since the relevant scales are far outside the horizon, $k \ll 
H$.
However, the condition that the relative entropy Eq.(\ref{adi}) vanishes, when 
$c_a^2=c_s^2$, would be strictly valid only for the instant that it is imposed. 
Thereby, although is therefore some arbitrariness in the choice of initial values,
the late evolution is not affected by the choice of these values as long 
as they are inside some reasonable region. We use, for all models, the 
initial values
\be
\delta  =  \frac{3}{4}(1+w)\delta_r, \quad
\theta  =  \theta_r, \quad
\sigma  =  0.
\ee   
The two first initial conditions are derived assuming that $c_s^2=c_a^2$ 
and that the relative entropy between dark energy and radiation together 
with its first derivatives vanish. The third condition says that we set 
the anisotropic stress to zero at very large
scales at an early time. Only then shear is not generated when $c_{vis}^2=0$. 
Thus fluids with vanishing viscous parameter are perfect. Note however 
that when the parameter is allowed to evolve with time, one can 
have $c_{vis}^2 = 0$ for an imperfect fluid at some stage.

\subsection{Dark Energy models with $-1<w<0$}

When $w$ is negative but larger than $-1$, the evolution of perturbations is 
determined by the two sound speeds of dark energy. 
The evolution has been analyzed in Ref.  \cite{Weller:2003hw} but without 
including the anisotropic stress. 

\begin{figure}
\begin{center}
\includegraphics[width=0.47\textwidth]{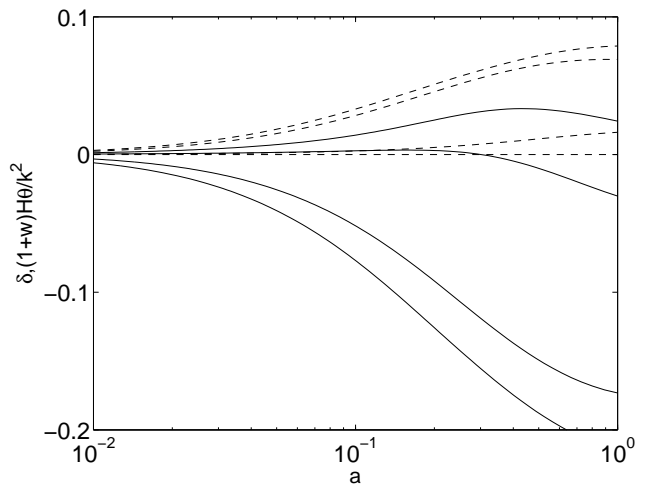}  
\caption{\label{perta} Late evolution of the dark energy density perturbation 
and velocity potential for $k=1.3\cdot 10^{-4}$ Mpc$^{-1}$ 
when $w=-0.8$. Solid lines from top to bottom correspond to $\delta$,
and dashed lines from bottom to top correspond to $(1+w)H\theta/k^2$
when ($c_{s}^2$, $c_{vis}^2$) = (0,0), (0.6,0), (0,0.6), (0.6,0.6).
The effect of $c_{vis}^2$ is to damp density perturbations,
which in the synchronous gauge is seen as a consequence of
enhancing the velocity perturbations.} 
\end{center}
\end{figure}

The metric perturbation $h$ is now a source in Eq. (\ref{deltaevol}), 
which tends to draw dark energy into 
overdensities of cold dark matter. However, for large scales the 
source due to velocity perturbations is proportional to 
$-(1+w)(c^2_s-w)\theta/k^2$, and 
this term can dominate the metric source term and drive $\delta$ to 
smaller values. In fact $\delta$ drops below zero when evaluated in the 
synchronous gauge \footnote{More accurately, we evaluate the transfer
functions of the perturbations. A negative value for the transfer function 
indicates that a perturbation variable acquires the opposite sign to its 
initial value.}. This happens especially for large sound speeds, since 
then both the friction term in Eq. (\ref{thetaevol}) and the source 
term in 
Eq. (\ref{deltaevol}) are larger (see FIG. \ref{perta}). 
Thus $\delta$ gets smaller when dark energy begins to dominate. The ISW 
effect is enhanced when one increases the sound speed squared. 
\begin{figure}
\begin{center}
\includegraphics[width=0.47\textwidth]{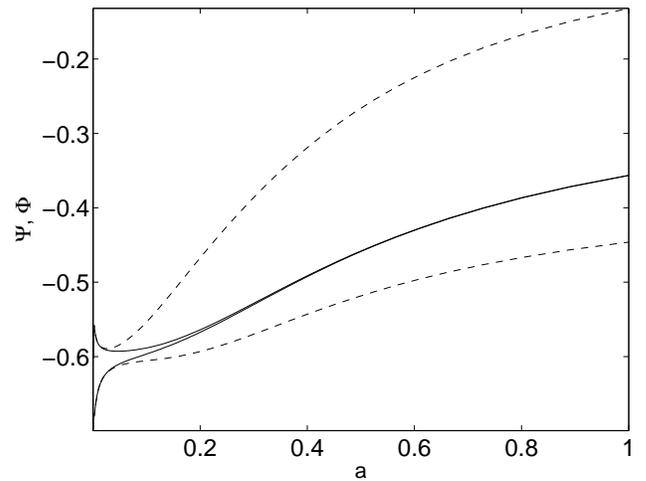}  
\caption{\label{gravis1} Late evolution of the gravitational
potentials at large scales ($k=1.3\cdot 10^{-4}$ Mpc$^{-1}$) 
when $w=-0.8$ and $c_s^2=0$. Solid lines are for the case of
perfect dark energy and dashed for the imperfect case with
$c_{vis}^2=1.0$ The upper lines are $\psi$, the lower lines are
$\phi$.} 
\end{center}
\end{figure}

The effect of the anisotropic stress is also to wash out overdensities. This 
is because the metric part of the source term in 
Eq. (\ref{sigmaevol}) turns out negative, and it dominates over the 
velocity term. Thus $\sigma$ is driven to negative values, and in 
Eq. (\ref{thetaevol}) it will act to increase the growth of $\theta$. 
This is similar to free-streaming of neutrinos, although for them
the effect is relevant at smaller scales. Since now $c_a^2<0$, the
source term $\sim H^2\theta/k^2$ in Eq. (\ref{deltaevol}) inhibits structure
growth at large scales. Therefore, as the dark energy becomes dominant,
the overall density structure is smaller when $c_{vis}^2$ is larger,
and the ISW effect is amplified. 

It is illuminating to describe the same thing also in terms of 
the Newtonian gauge perturbations. This gauge is defined by the line 
element
\be
ds^2 = a^2(\tau)\left[-(1+2\phi)d\tau^2 + (1-2\psi)dx^idx_i\right].
\ee
Here $\tau$ is the conformal time.
We remind that the ISW effect stems from the time variation of the metric 
fluctuations,
\be \label{isw}
C^{ISW}_\ell \propto \int\frac{dk}{k}\left[\int_0^{\tau_{LSS}} d\tau 
(\dot{\phi} + \dot{\psi})j_\ell(k\tau)\right]^2,
\ee
where $\tau_{LSS}$ is the conformal distance to the last scattering 
surface and $j_\ell$ the $\ell$'th spherical Bessel function. The ISW 
effect occurs 
because photons can gain energy as they travel trough time-varying 
gravitational wells. These wells are in turn caused by matter, since
\be \label{phi}
-k^2\phi = 4\pi G a^2\rho\left[\delta + 3\frac{H}{k^2}(1+w)\theta\right]_{|T}.
\ee
We have indicated with the subscript $ _{|T}$ that in the left hand side 
variables refer to all matter present, and not just dark energy. Note also 
that the term in square brackets is gauge-invariant. Thus, evaluated in 
any 
frame, it equals $\delta_T$, the overdensity of energy seen in the 
comoving frame. During matter domination, $\delta_T$ grows in such a way 
that the gravitational potentials stay constant. It is then clear 
that as dark energy begins to take over, the gravitational potential 
$|\phi|$ begins to decay. Contrary to expectations from FIG. \ref{perta},
this decay is not more efficient at large scales when there is shear, as 
shown in FIG. \ref{gravis1}. This is because the dark energy shear 
influences gravitational wells in such a way that the growth of matter 
perturbations does not slow down as much as in a perfect universe.  

However, there is an important twist to the story. This is seen in the 
FIG. \ref{gravis1}, where the evolution of the potentials $\phi$ and 
$\psi$ is plotted at very large scales. At an early time the potentials 
are unequal because of the free streaming of radiation. However, our 
attention is now on the late evolution of the potentials. Due to dark 
energy, the potentials can re-depart from each other at smaller redshifts. 
This can happen only when $c_{vis}^2 \neq 0$, since
\be \label{psi}
\psi = \phi - 12 \pi G a^2(1+w)\rho \sigma_{|T},
\ee
i.e. shear is the difference between the depth of matter-induced gravity 
well and the amount of spatial curvature. Since $\sigma$ is 
gauge-invariant, and we found that it becomes negative for dark energy,
we can see that shear perturbation drives $|\psi|$ to vanish more 
efficiently. Thereby we find that the effect of shear on Eq.(\ref{phi}) 
only partly compensates for the effect on $\psi$ from 
Eq.(\ref{psi}), and thus the overall ISW from Eq.(\ref{isw}) will be 
amplified when dark energy perturbations tend to smooth as in FIG. \ref{perta}.

\begin{figure}
\begin{center}
\includegraphics[width=0.47\textwidth]{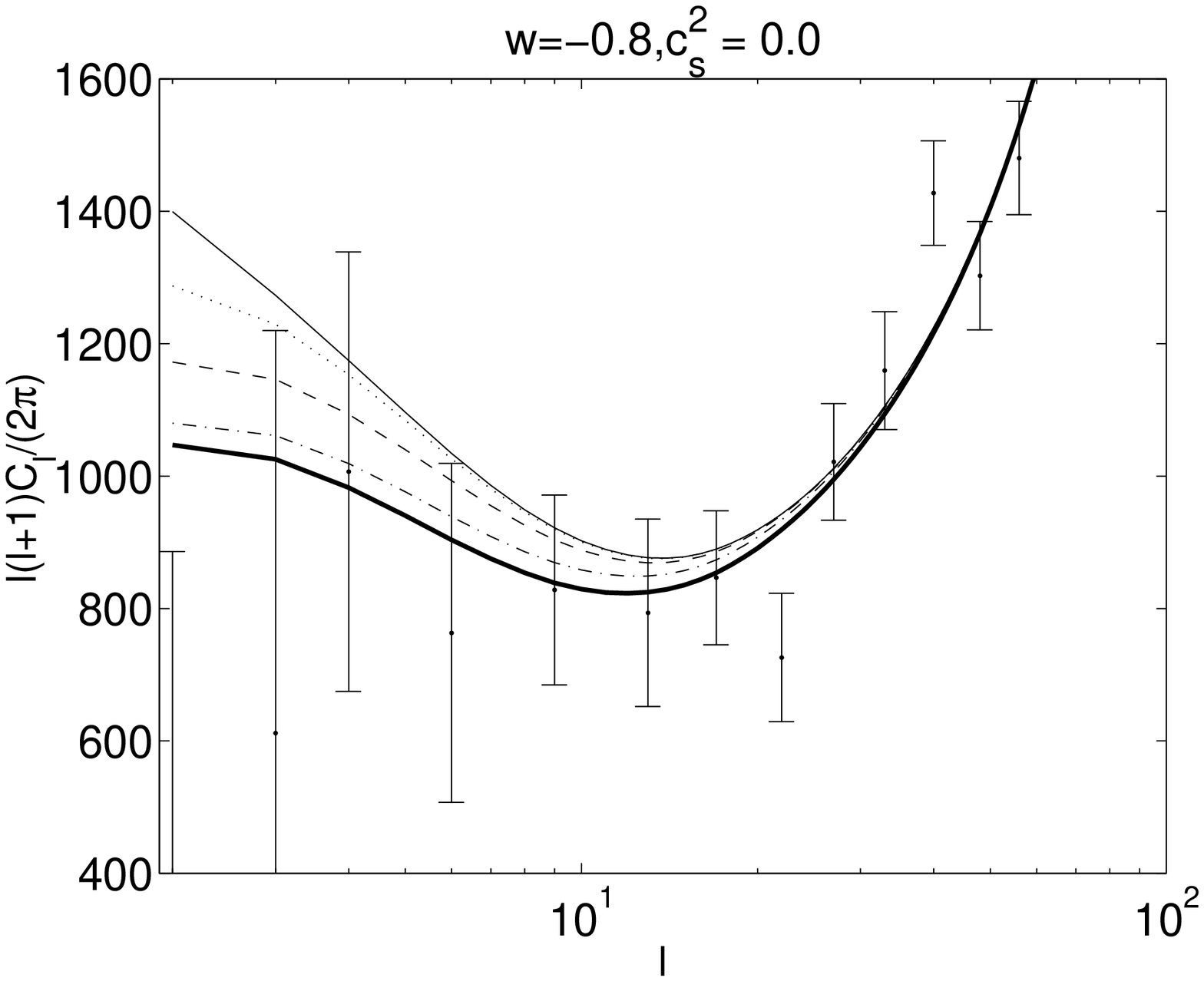}
\includegraphics[width=0.47\textwidth]{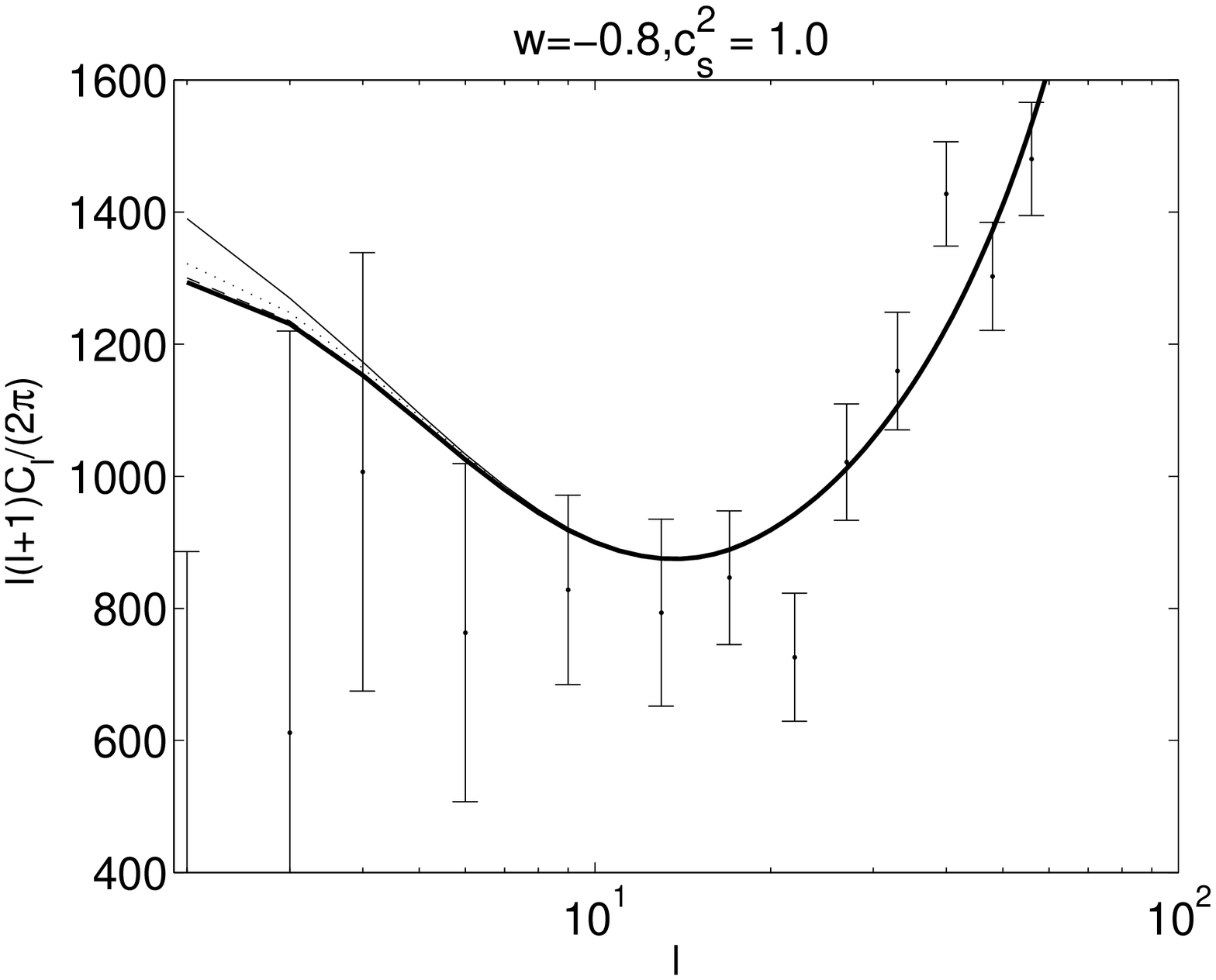}
\caption{\label{sosspic1} The CMBR anisotropies for $w=-0.8$. In the 
upper
panel $c_s^2=0$ and in the lower panel $c_s^2=1.0$. The ISW
contribution
increases with the parameter $c^2_{vis}$: thick lines are for
$c^2_{vis}=0$, dash-dotted for $c^2_{vis}=0.001$, dashed for
$c^2_{vis}=0.01$, dotted for $c^2_{vis}=0.1$ and the solid lines
for
$c^2_{vis}=1.0$.}
\end{center}
\end{figure}

In FIG. \ref{sosspic1} we show the large angular scales of the CMBR 
spectrum when $w=-0.8$ and the two other parameters are varied. The upper
panel depicts the case where the sound speed of dark energy vanishes. Then
the pressure perturbation vanishes and the clustering
of dark energy is inhibited only by the free-streaming effect of shear 
viscosity. Therefore the large scale power of the CMBR is increased by 
increasing $c_{vis}^2$. In the lower panel 
$c_s^2=1$. Then dark energy is almost smooth (except at the largest 
scales) even without anisotropic stress, and thus we see a smaller effect 
when $c_{vis}^2$ is increased. 

When $c_{vis}^2<0$ the metric 
and the fluid sources drive the perturbations in the same direction, 
resulting in explosive growth. Since this would spoil the evolution except 
when $c_{vis}^2$ is tuned to infinitesimal negative values, we will not 
consider such a case here. 

\subsection{Phantom Dark Energy models with $w<-1$}

\begin{figure}
\begin{center}
\includegraphics[width=0.47\textwidth]{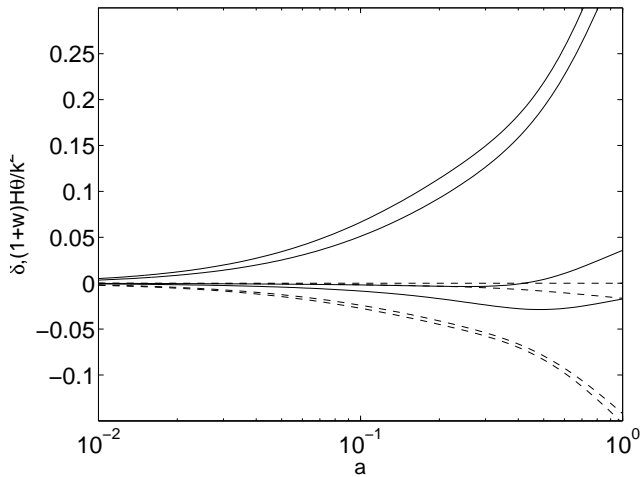}  
\caption{\label{pertb}
Late evolution of the dark energy density perturbation and
and the velocity potential for $k=1.3 \cdot 10^{-4}$ Mpc$^{-1}$
when $w=-1.2$. Solid lines from bottom to top correspond to $\delta$, the 
dashed lines from top to bottom correspond to $(1+w)H\theta/k^2$ when
($c_s^2$, $c_{vis}^2$) = (0,0), (0.6,0), (0,-0.6), (0.6,-0.6).
The effect of $c_{vis}^2$ is to increase clustering,
which in the synchronous gauge is seen as a consequence of
enhancing the velocity perturbations.}
\end{center}
\end{figure}

When the dark energy equation of state is less than $-1$, the effect of 
both the sound speed and of the viscosity parameter are the opposite to 
the previous case. Now 
the source term in Eq. (\ref{deltaevol}) has its sign reversed, and because 
of that dark energy falls out from the overdensities. Similarly, the 
velocity potential acts now as a source for the 
overdensities. Therefore increasing the 
sound speed will drive dark energy to cluster more efficiently. Now the 
effect of $c_{vis}^2>0$ is with the same sign of those of the metric 
sources, and therefore we must consider negative values for this 
parameter. Then, if we increase the parameter $c_{vis}^2/(1+w)$, the dark 
energy perturbations are growing more efficiently, as shown in FIG. 
\ref{pertb}. This is because $\sigma$ is negative, just like in the 
previous case, and again tends to enhance the velocity potential. The 
crucial difference in the perturbation evolution for imperfect
dark energy here as compared to the imperfect 
$w>-1$ case is that more shear in the perturbations will 
result in more clumpy structure in the density of
phantom dark energy.   
\begin{figure}
\begin{center}
\includegraphics[width=0.47\textwidth]{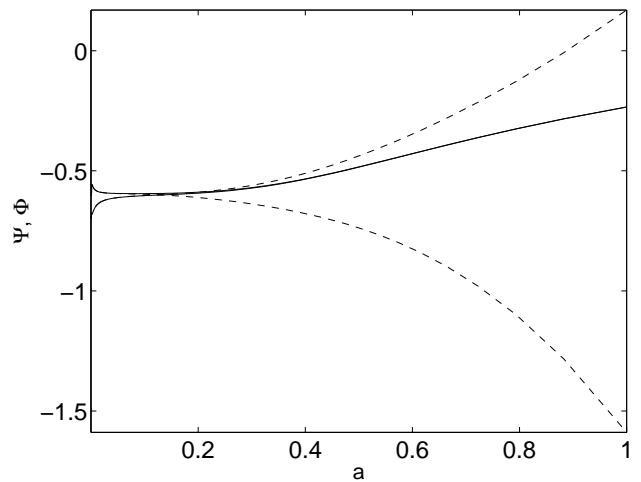}  
\caption{\label{gravis2} Late evolution of the gravitational
potentials at large scales( $k=1.3\cdot 10^{-4}$ Mpc$^{-1}$) 
when $w=-1.2$ and $c_s^2=0$. Solid lines are for the case of
perfect dark energy and, dashed for the imperfect case with
$c_{vis}^2=-1.0$. The upper lines are $\phi$, the lower lines are $\psi$.} 
\end{center}
\end{figure}

One can again consider the ISW in the terms of the Newtonian gauge 
potentials, Eq.(\ref{isw}). The effect is not directly seen from 
the behaviour of $\delta$ in FIG. \ref{pertb}, partly because of the 
different gauge and partly because the anisotropic stress induces a 
compensation on the other gravitional potential and thereby influences 
also the matter perturbation. This is shown in FIG. \ref{gravis2}.
The anticipated simple result (that the decay of the gravitational 
potentials is reduced since $\delta$ is enchanced when there is more 
shear) again holds for the sum of the gravitational potentials, but 
considering $\phi$ or $\psi$ separately reveals the intricacy of the 
fluctuation dynamics due to anisotropic stress.
Again evolution of $\phi$ implies, through Eq.(\ref{phi}), that the 
influence of $\sigma$ to dark matter is the opposite from dark energy, but on the
other hand, evolution of the spatial curvature $\psi$ implies that the 
sum $\phi+\psi$ behaves according to the dominating component. Now 
the gravitational well $\psi$ grows deeper, because the contribution from 
shear in the phantom fluid in Eq.(\ref{psi}) comes with a minus sign.

\begin{figure}
\begin{center}
\includegraphics[width=0.47\textwidth]{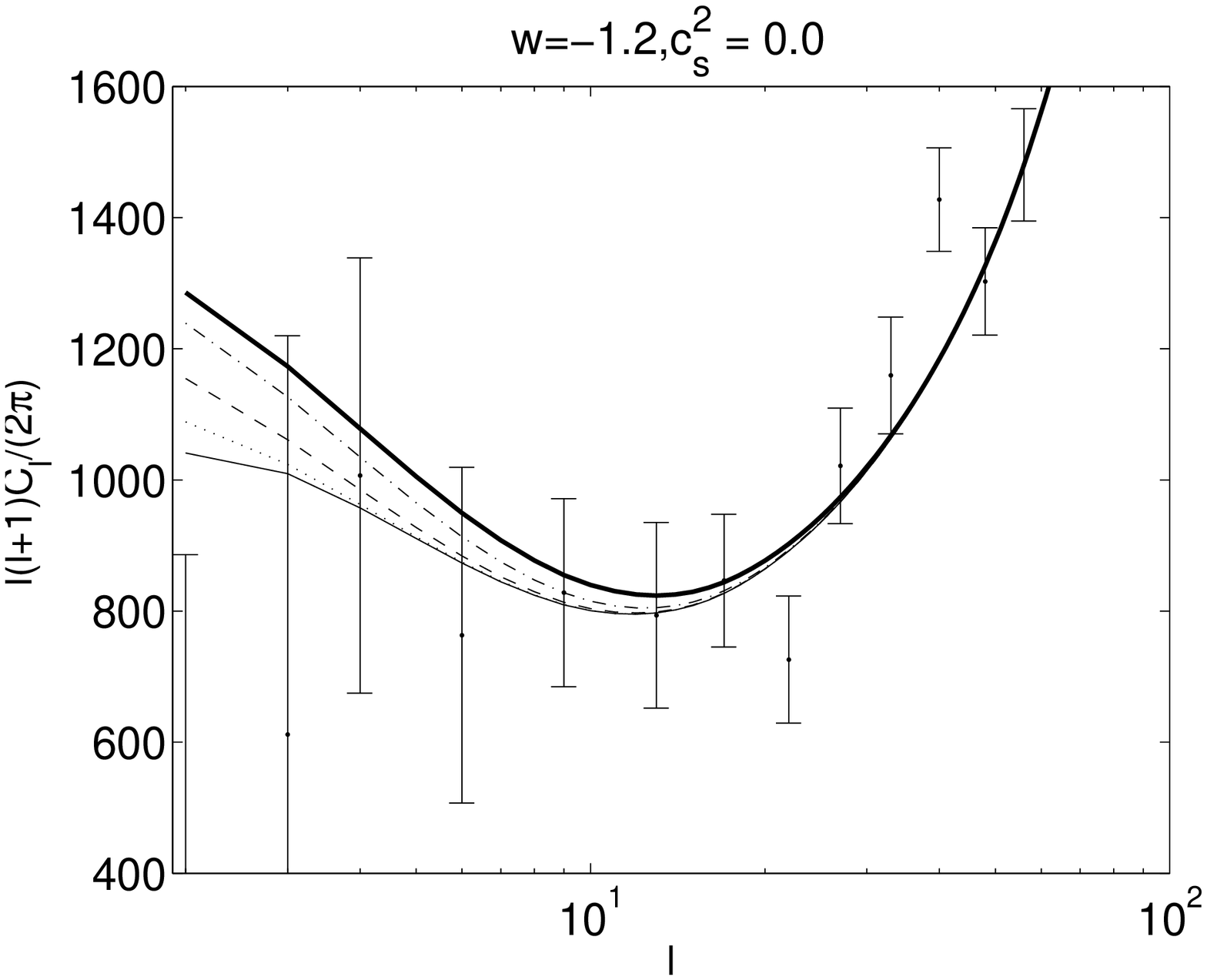}
\includegraphics[width=0.47\textwidth]{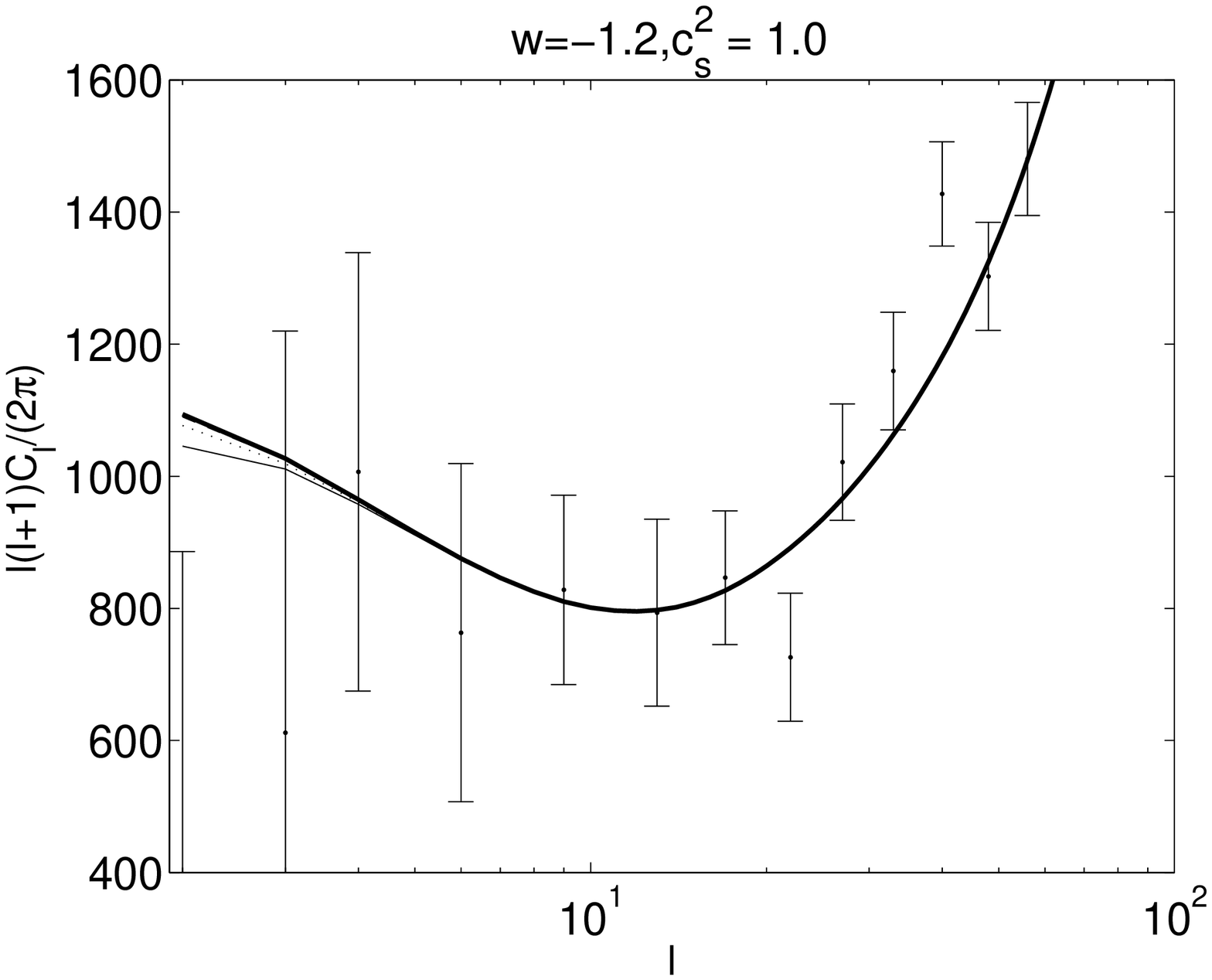}
\caption{\label{sosspic2}
The CMBR anisotropies for $w=-1.2$. In the upper
panel $c_s^2=0$ and in the lower panel $c_s^2=1.0$. The ISW
contribution
decreases with the parameter $c^2_{vis}$: thick lines are for
$c^2_{vis}=0$, dash-dotted lines for $c^2_{vis}=0.001$, dashed lines for
$c^2_{vis}=0.01$, dotted lines for $c^2_{vis}=0.1$ and the solid lines
for $c^2_{vis}=1.0$.}
\end{center}
\end{figure}

In FIG. \ref{sosspic2} we have plotted the large angular scales of the CMBR 
spectrum when $w=-1.2$ and the two other parameters are varied. The upper
panel depicts the case that the sound speed of dark energy vanishes. Then 
the ISW effect without anisotropic stress is large since
dark energy perturbations are nearly washed out. Consequently, the large scale
power of CMBR is decreased as $|c_{vis}^2|$ is increased, since 
the ''anti-viscosity'' will then amplify perturbations. In the lower panel 
$c_s^2=1$. There the effect of $c_s^2$ already dominates, and 
we see a smaller difference when $c_{vis}^2/(1+w)$ is increased. 

\subsection{Dark Energy models with $c_s^2<0$}

For perfect dark energy models without shear, the case $c_s^2<0$
leads to explosive growth of perturbations. This is  
analogous to the behaviour of a simple wave, which has a solution 
$\sim e^{-i(k/c_s) t + i\bar{k}\cdot\bar{x}}$, diverging
when the
sound velocity is imaginary. It is not clear, however, how  
useful the analogy to the sound speed of a simple plane wave is to the 
interpretation of the variable defined by Eq. (\ref{cs}). For instance, in 
the modified gravity context  
 \cite{Koivisto:2004ne} this formal definition does not 
describe propagation of waves in any physical matter. A priori 
one should not discard the possibility $c_s^2<0$ without careful 
deliberation. 
In fact, given a fluid with negative equation of state,
one would expect, from Eq. (\ref{ca}), also a negative sound speed squared. 
To get rid of this feature, extra degrees of freedom must be assumed to 
exist in such a way that the variable defined by Eq. (\ref{cs}) turns out 
positive.  

\begin{figure}
\begin{center}
\includegraphics[width=0.47\textwidth]{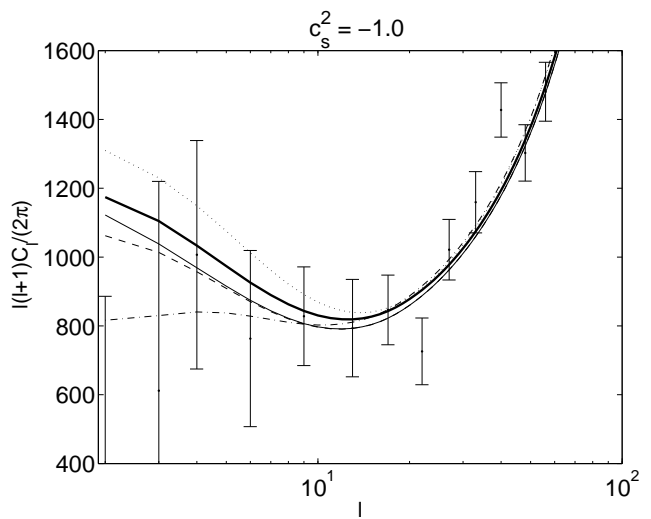}
\caption{\label{sosspic3} The CMBR spectra for 
$c_s^2=-1.0$. The thick line is for $w=-1$ (unperturbed
dark energy). 
The solid
line is for $w=-0.8$ and $c_{vis}^2=0.5$ and the
dash-dotted line for $w=-0.8$ and $c_{vis}^2=1.0$
The dotted line is for $w=-1.2$ and $c_{vis}^2=0.5$ and the
dashed line for $w=-1.2$ and $c_{vis}^2=1.0$. }
\end{center}
\end{figure}

When the generation of shear in the fluid is taken into account, the
perturbation growth for $c_s^2<0$ can be stabilized. This is because shear 
is sourced by the perturbations, and in turn the shear will inhibit 
clustering. Here it is possible to choose the parameters in such a way 
that the dark energy perturbation grows steadily at late times. So the ISW 
effect comes with the opposite sign from the Sachs-Wolfe effect, which 
leaves its imprint in the CMBR earlier. These effects cancel each other 
and thus the large scale
power in the CMBR spectrum is reduced, in accordance with the 
measured low quadrupole. We show in FIG. \ref{sosspic3} such a case 
together with various other choices for the other two parameters when
$c_s^2=-1$.  

\subsection{A summary}

We summarize the features of dark energy perturbations in different 
parameter regions in Table \ref{tab}. A half of the parameter space is 
excluded because 
divergent behaviour occurs, and much of the remaining parameter space 
is degenerated. Even when restricting to simplest case where the 
parameters are kept constant, it seems clear 
that present observational data allows large variety of interesting models 
with 
non-vanishing shear, $c_{vis}^2 \neq 0$. In some parameter regions 
of Table \ref{tab} new features appear at observable scales.
\begin{table}
\begin{tabular}{|c|c|c|c|c|}
\hline
$w$  & $c_s^2$ & $c_{vis}^2<0$ & $c_{vis}^2 = 0$ &  $c_{vis}^2>0$ \\
\hline
$ > -1 $ & $ > 0$  &  diverges    & canonic scalar field   &  
$\searrow \phantom{\dag}$  (FIG. \ref{sosspic1})  \\
\cline{2-5}
$   $ & $ < 0$  &  diverges   &     diverges    &  $\nearrow$ \dag (FIG. \ref{sosspic3})         \\
\hline
$<-1  $ & $ > 0$  &  $\nearrow \phantom{\dag}$ (FIG. \ref{sosspic2})  &  phantom scalar field &    diverges        \\
\cline{2-5}
$     $ & $ < 0$  &  $\searrow$ $\dag$ (FIG. \ref{sosspic3})   &     diverges          &    diverges       \\
\hline
\end{tabular} 
\caption{\label{tab} Summary of different parameter regions for dark 
energy fluids. We have indicated with $\nearrow$ the cases where superhorizon
perturbations are increased as $|c_{vis}^2|$ is increased, and with
$\searrow$ the cases where superhorizon perturbations in dark energy are
smoothened as $c_{vis}^2$ is increased. We indicate by $\dag$  
that the shear perturbation influences significantly also the small scale 
perturbations. 
}
 
\end{table}

\begin{figure}
\begin{center}
\includegraphics[width=0.47\textwidth]{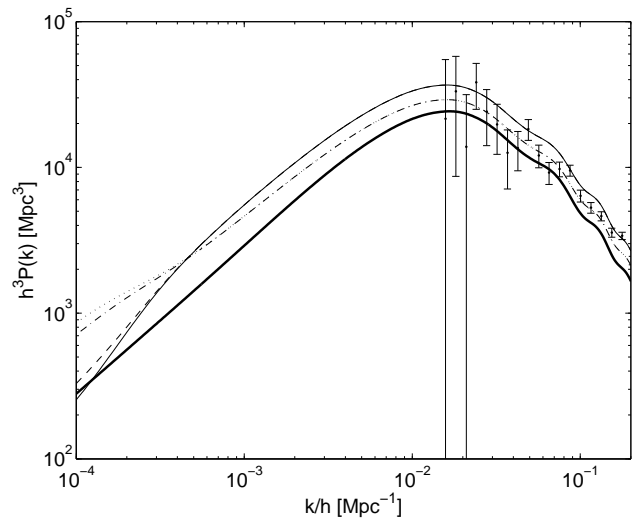}
\caption{\label{matterpic1} The total matter power spectra 
when $c^2_{vis}=1.0$. The thick line is for $w=-0.8$ and 
$c_s^2=1.0$. The dash-dotted line for $w=-0.8$ and $c_s^2=0$, the
dotted line for $w=-0.8$ and $c_s^2=1$. The dashed line
is for $w=-1.2$ and $c_s^2=0$, the solid line for $w=-1.2$
and $c_s^2=1$.}
\end{center}
\end{figure}

In FIG. \ref{matterpic1} we plot the matter power spectra including cold 
dark matter and baryons 
for various parameter choices. When $c_{vis}^2$ is varied,
the effect occurs only at scales much larger than what current 
observations are able to probe. In the spectrum of the total density 
perturbation
one would see more pronounced features at scales more tantalizingly 
near the current limits of observations. However, there is no way to
directly measure the dark energy density perturbation, and therefore
we have plotted only the power spectrum of non-relativistic matter.

In FIG. \ref{gravis1} and 
FIG. \ref{gravis2} it was seen that the shear changes the Newtonian
gravitational potentials significantly. 
Thus one might hope to find a way to study whether effects from an anisotropic stress
could be measured by using for example the cross correlation of the ISW signal 
and the large scale structure observations or gravitational lensing experiments. 
However, one should keep in mind that we have considered perturbations at 
vast scales. We have found that 
that fluctuations in an imperfect as well as in a perfect fluid with a 
constant equation of state $w<-1/3$ are confined to superhorizon scales. 
This is except for special occasions, in particular when the
parameter $c_s^2$ is negative or when the perturbations behave
pathologically due to wrong sign of the viscous parameter.  
For the largest scales, the viscous parameter determines the evolution
of the perturbations. It is clear from FIG. (\ref{perta}) and 
FIG. (\ref{pertb}) that the variation of the sound speed $c_s^2$
has much less effect on the evolution of perturbations in the limit
$k \rightarrow 0$. However, the parameter $c_s^2$ sets the
scale at which the fluctuations in dark energy become negligible. For
smaller $c_s^2$, there are fluctuations at smaller wavelengths. 
Therefore the shear would be best seen when the $c_s^2$ is nearly
zero or even negative. 

The main impact of dark energy anisotropic stress on observations 
seems to be the modification of the CMBR at very large scales, from which 
it would be very difficult to unambiguously detect. However, this changes 
when one considers the perhaps 
better physically motivated situation where the parameters $w$, $c_s^2$ 
and $c_{vis}^2$ are allowed to evolve in time. 
   
\section{Imperfect unified models}

The Chaplygin gas  \cite{Kamenshchik:2001cp} is a prototype
of a unified model of dark matter and dark energy  \cite{Bilic:2001cg}.
In such models a single energy component accounts for both 
the dark matter and dark energy. Thus this component must
resemble cold dark matter in the earlier universe, whereas it should exhibit 
large negative pressure nowadays.
These models are, however, problematic because
of the suppression of structure formation by the adiabatic pressure
perturbations  \cite{Carturan:2002si,Sandvik:2002jz,Amendola:2003bz}. 
A solution for this problem has been based on  
the observation that due to entropy, the sound speed is not 
necessarily the adiabatic one  
\cite{Reis:2003mw,Koivisto:2004ne,Zimdahl:2005ir}. In the 
so called silent quartessence model entropy perturbations cancel the 
effect of the adiabatic sound speed  \cite{Amendola:2005rk}. 
 
The modified polytropic Cardassian expansion  \cite{Freese:2002gv} (MPC) in 
the fluid 
interpretation  \cite{Gondolo:2002fh} provides a general parameterization
which encompasses a wide variety of unified models. For the MPC
case, one can write the energy density  
as a function of the scale factor as
\be \label{quart}
\rho = [A a^{3 q(\nu-1)} + B a^{-3q}]^{\frac{1}{q}}.
\ee
The exponents $q$, $\nu$ are given as parameters, and the
constants $A$, $B$ have the appropriate mass dimension. This 
parameterization is equivalent to 
the New Generalized Chaplygin gas  \cite{Zhang:2004gc}.
When $\nu=2$, one gets the 
Generalized Chaplygin gas  \cite{Bento:2002ps} where $q$ can vary,
and setting further $q=1$, one is left with the original Chaplygin gas 
 \cite{Kamenshchik:2001cp}.
On the other hand, when the parameter $\nu$ can get any values and $q$ is 
set to $q=2$, 
the Variable Chaplygin gas  \cite{Guo:2005qy} is recovered. Finally, 
when $q=1$ and $\nu$ is arbitrary, one has the simplest version of the Cardassian 
expansion which reproduces the background expansion of a universe with 
standard CDM and dark energy with $w=\nu-2$  \cite{Freese:2002sq}.

We will consider here effects of shear in models where the unified dark density 
is defined by Eq. (\ref{quart}). In the above references, various theoretical 
origins for such density ansatzes have been proposed, but we will not
consider here whether these necessitate incorporation of 
shear in the linear order in cosmology.  
Previously Cardassian expansion has been studied in the modified
gravity context  \cite{Koivisto:2004ne}, and there it was shown that 
an effective anisotropic stress can appear in the late
universe. With certain assumptions for the modified gravity, the cold dark 
matter density perturbation 
generates a shear perturbation algebraically
determined from the density and velocity fields
of the matter interestingly similarly to the case
motivated by the covariant generalization of the Navier-Stokes
equation (\ref{l-l}). Then the resulting matter power spectrum 
is in better accordance with observations than in the
standard adiabatic case, but as the anisotropic stress
affects the gravitational potentials and thus enhances the
ISW effect, the CMBR spectrum will then restrict the allowed
parameter space stringently. However, we will here consider
the anisotropically stressed fluid as parameterized by
Eq.(\ref{sigmaevol}). As expected, our results 
will be different from the modified gravity
approach of Ref. \cite{Koivisto:2004ne}.

The equation of state for the unified fluid described by Eq.(\ref{quart}) 
is
\be
w = \frac{\nu A a^{3q(\nu-1)}}{A a^{3 q(\nu-1)} + B a^{-3q}},
\ee
and it follows that
\be
c_a^2 = \frac{w\left[1-\nu q+w(1-q)\right]}{1+w}.
\ee
When both $q$ and $\nu$ are equal to one, the model is equivalent
to cold dark matter and a cosmological constant. When either
$q$ or $\nu$ is smaller than one, $c_a^2$ will be negative
in the late universe, and when either $q$ or $\nu$ is greater
than one, the $c_s^2$ will stay positive and grow in the 
late universe. For instance, if $\nu=1$, the asymptotic value
is $c_s^2=1-q$. 

In the adiabatic case then $c_s = c_a$, but for the silent quartessence
one imposes a special condition on $c_s^2$, namely that the pressure 
perturbation vanishes in the synchronous gauge  
\cite{Reis:2003mw,Amendola:2005rk}. 
Here we consider the case
that all the three sound speeds are equal in magnitude, including the
viscosity parameter as it appears in 
Eq. (\ref{sigmaevol}). Thus $c^2_{vis}=|c_a^2|$ and $c_s^2=c_a^2$. This seems a 
natural generalization of the characteristics of better known cosmic fluids, 
i.e. neutrinos. Then, in the terminology employed here, the fluid is 
adiabatic but imperfect\footnote{The presence of anisotropic stress does 
not lead to generation of entropy at the linear order 
 \cite{Maartens:1997sh}.}. For comparison, we include also 
results for 
the silent quartessence, as an example of an entropic but perfect fluid. 
In addition, results for the adiabatic and perfect model are shown.

\begin{figure}
\begin{center}
\includegraphics[width=0.47\textwidth]{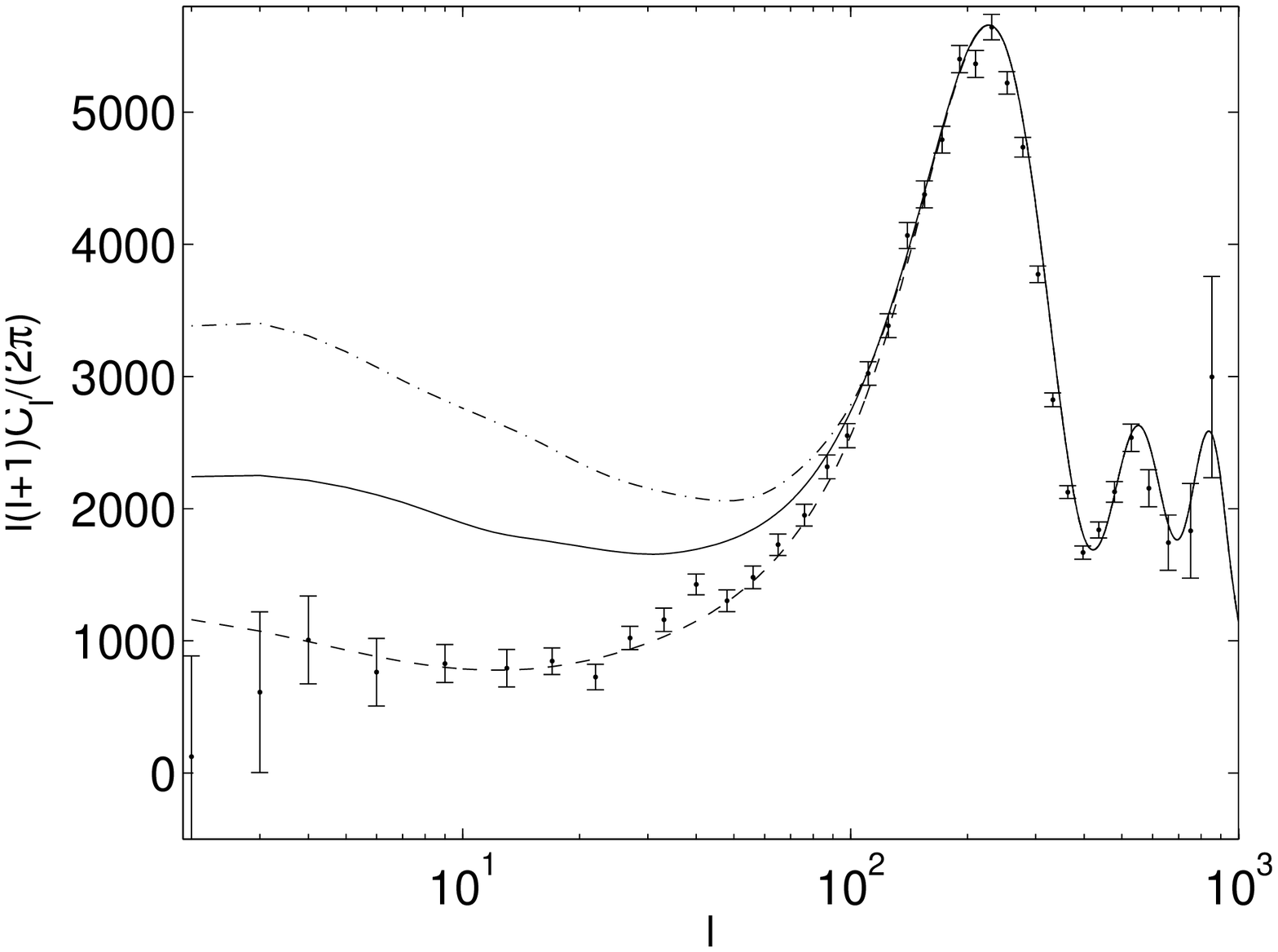}  
\includegraphics[width=0.47\textwidth]{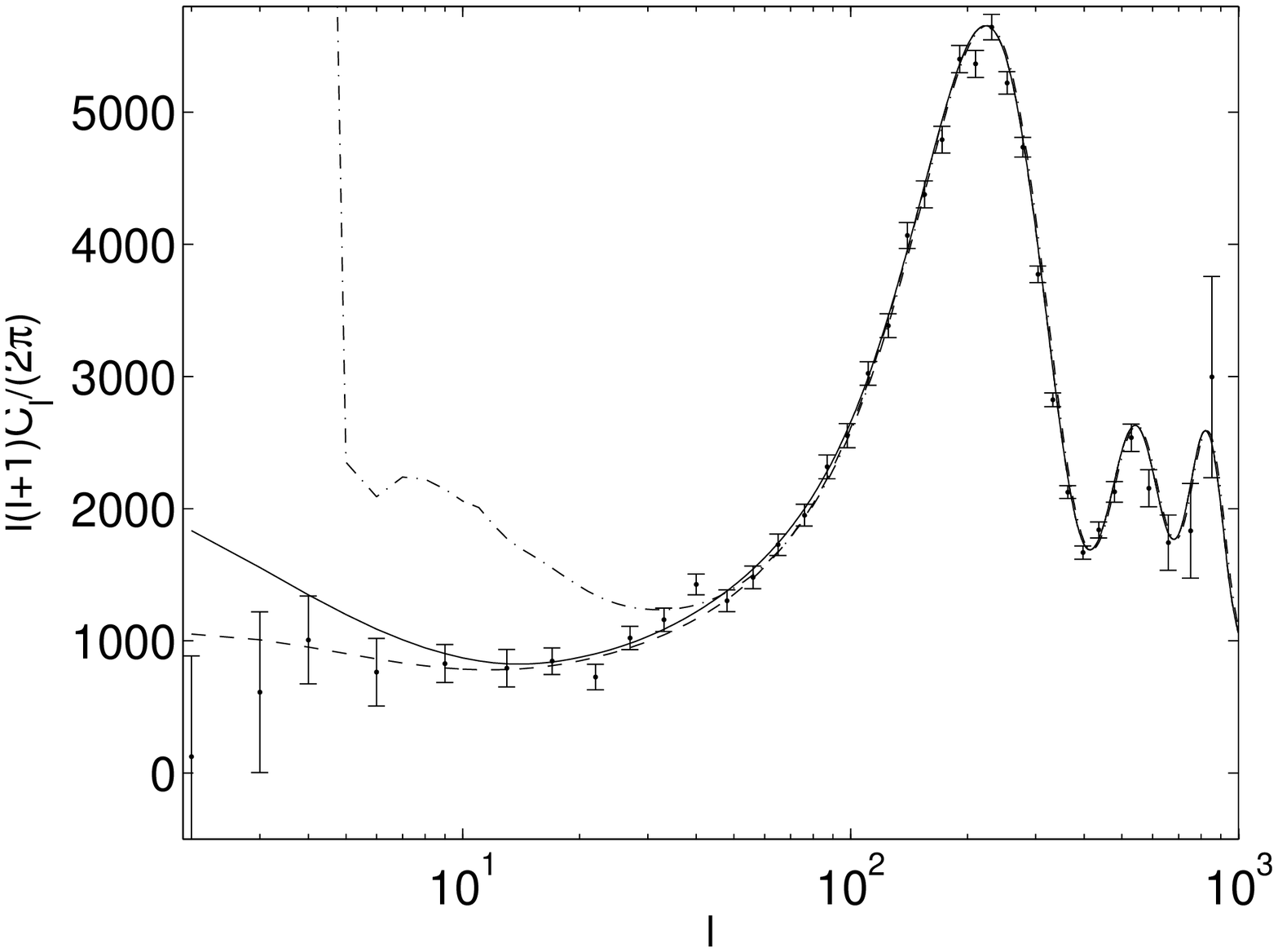}  
\caption{\label{chaps} The CMBR anisotropies for the MPC model
with $q=1.0$. The dash-dotted lines are for the adiabatic case, the solid
lines for the same model with the shear included, and the dashed lines 
correspond to the silent case. In the upper panel $\nu=1.1$. 
In the lower panel $\nu=0.9$ for the silent and the imperfect case.
The adiabatic model has $\nu=0.994$, which already gives disproportionately
large ISW effect.}
\end{center}
\end{figure}

\begin{figure}
\begin{center}
\includegraphics[width=0.47\textwidth]{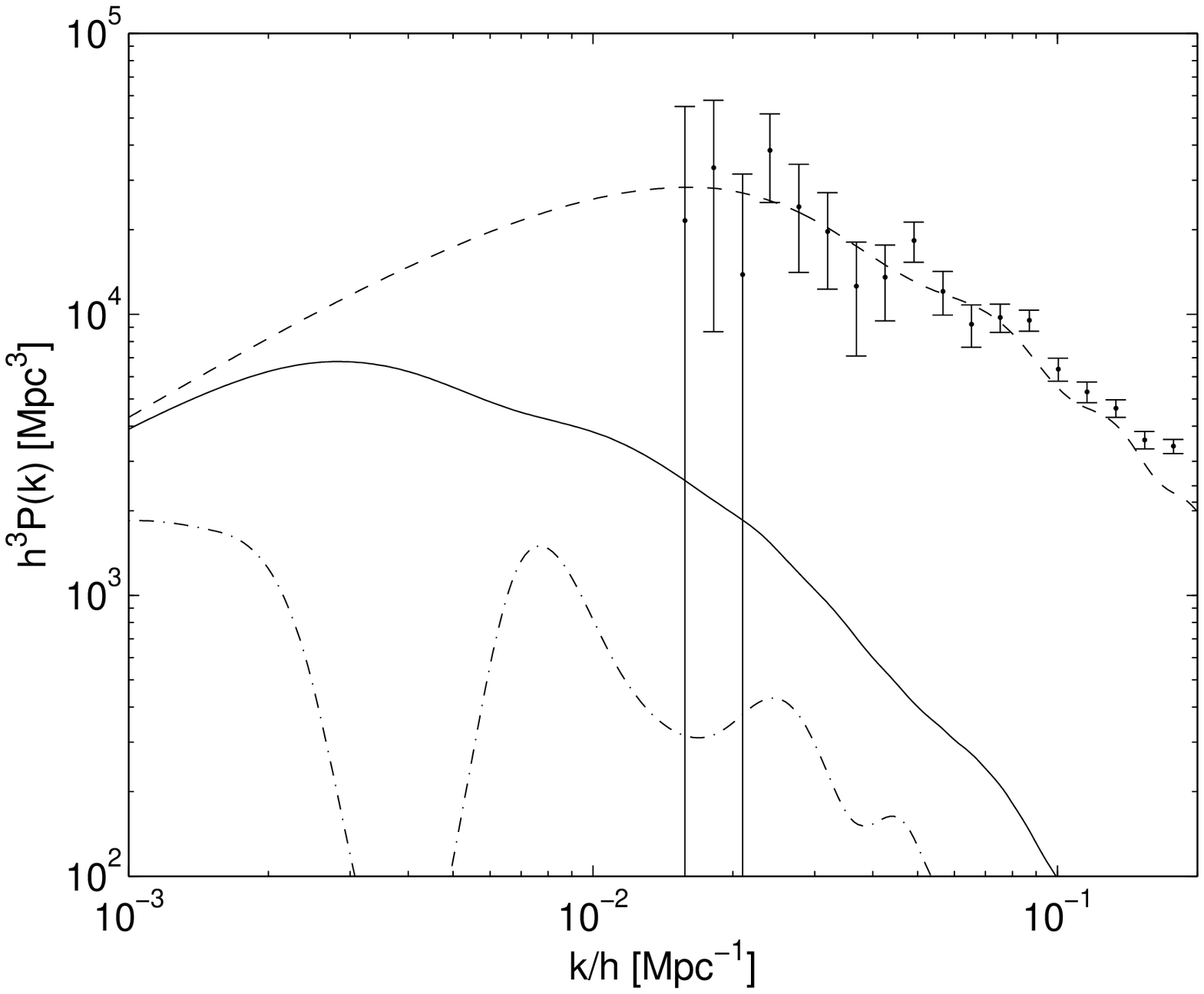}  
\includegraphics[width=0.47\textwidth]{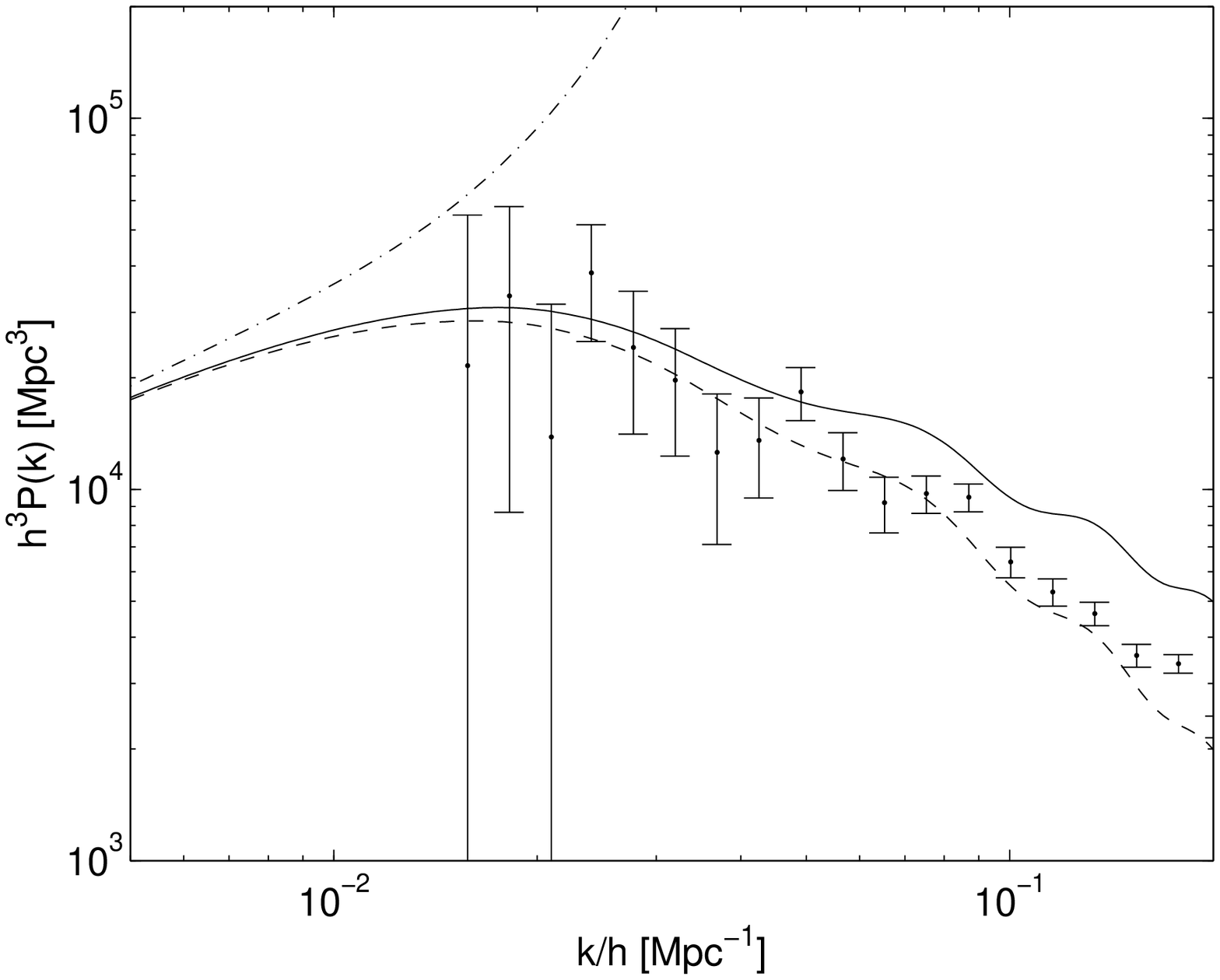}  
\caption{\label{mpschaps1} The total matter power spectra for
MPC model when $\nu=1.0$ In the upper panel $\nu=1.01$,
and in the lower panel $\nu=0.999$.}
\end{center}
\end{figure}

\begin{figure}
\begin{center}
\includegraphics[width=0.47\textwidth]{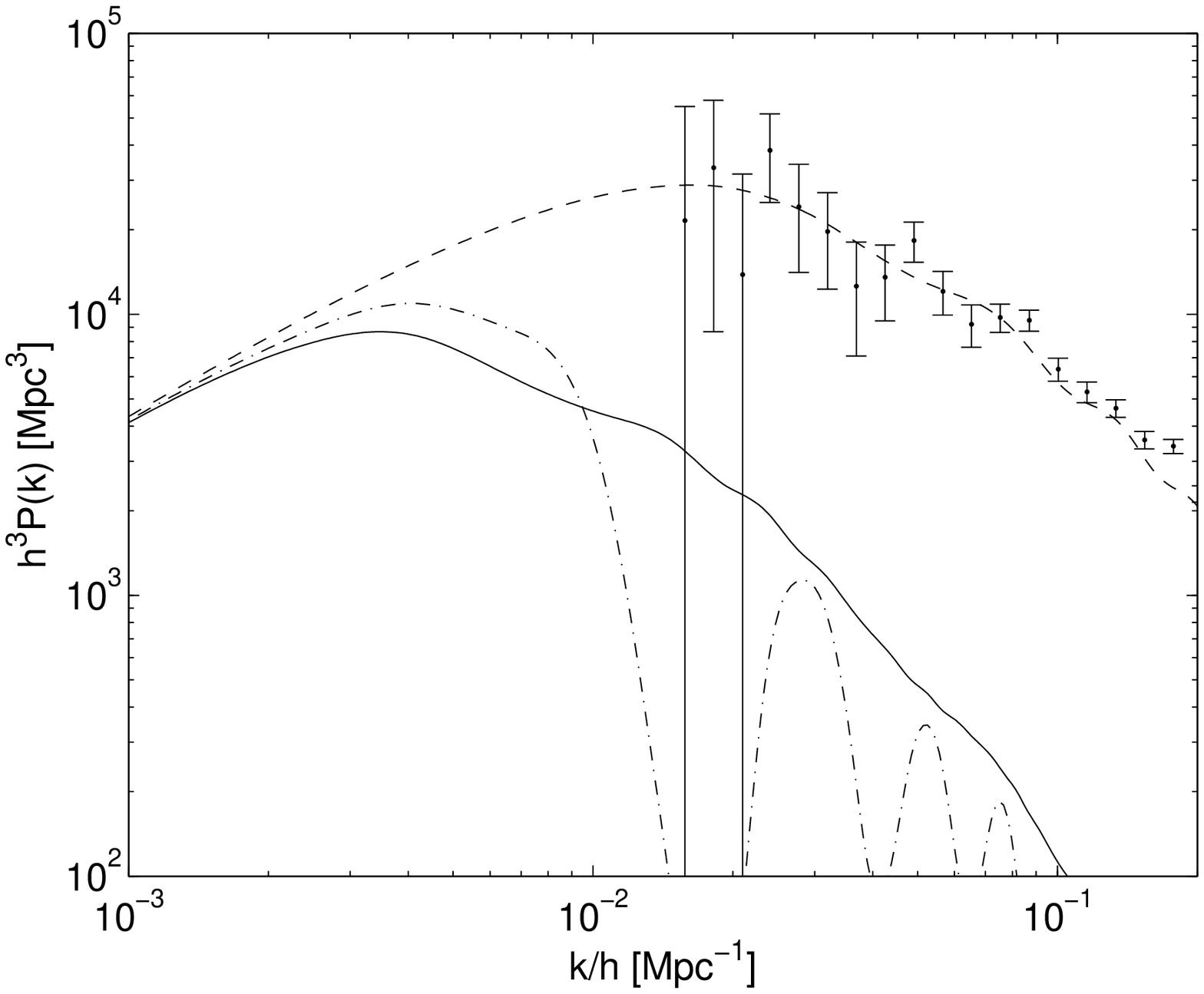}  
\includegraphics[width=0.47\textwidth]{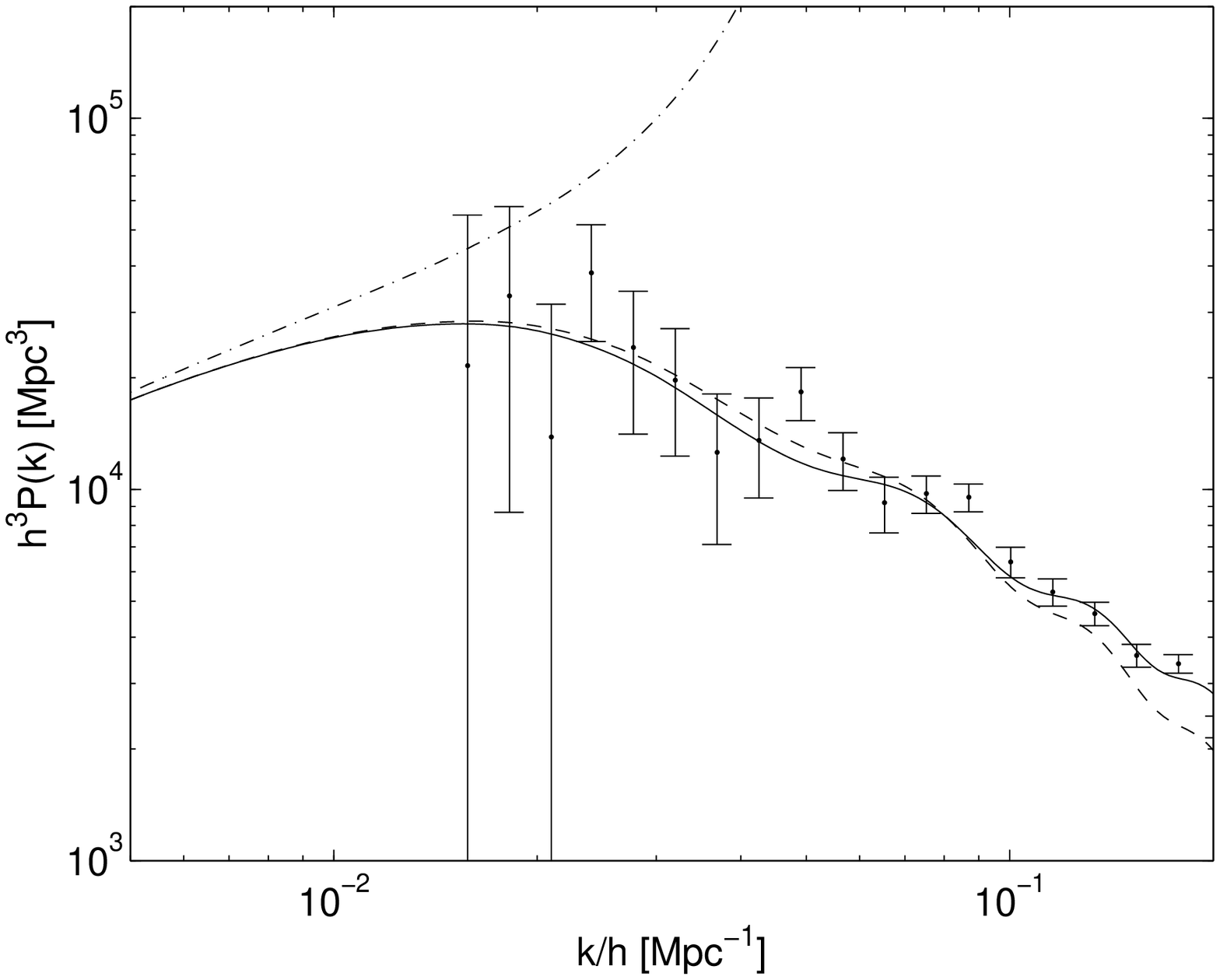}  
\caption{\label{mpschaps2} The total matter power spectra for
MPC model when $\nu=1.0$. In the upper panel $q=1.01$,
and in the lower panel $q=0.999$.
}
\end{center}
\end{figure}

We plot the results for the CMBR spectrum in FIG. \ref{chaps},
and for the matter power spectra in FIG. \ref{mpschaps1}
and FIG. \ref{mpschaps2}. Here the matter power spectra include
all the components in the energy budget, since one cannot distinguish 
dark matter from the unified model. In all the figures, 
the adiabatic case is shown with dash-dotted lines, the entropic 
case with dashed lines and the imperfect case with solid lines. 

The effect of shear is to stabilize the perturbations. When
$q$ or $\nu$ is greater than one, the adiabatic pressure tends
to drive
the density perturbation to oscillate. However, including the anisotropic 
stress will remove the oscillations, since the damping effect of shear
compensates steep gradients. The overall suppression
of growth is alleviated, but not as much as in the silent model.
When either $q$ or $\nu$ is smaller than one, the adiabatic
pressure would drive the density perturbation to a very fast growth.
However, as discussed in the previous section, the shear viscosity
eliminates this driving force. 

\section{Conclusions}

In this article we have investigated the effects of an anisotropic stress 
in the dark energy component on large scale structures. 

We have parameterized the dark energy component with three variables. The 
equation of state
determines the decay rate of dark energy, and the sound speed
characterizes the evolution of its fluctuations.
These two were treated as independent parameters, thus accounting 
for possible entropy in the fluid. In addition we allowed for shear 
viscosity in the linear order. We discussed the possibility to apply a 
Navier-Stokes type viscosity to determine the additional degree of freedom for dark energy
fluctuations, the amount of shear viscosity, but we adopted the 
parameterization utilizing a viscosity parameter $c_{vis}^2$, motivated by 
the fact that it seems to generalize the familiar and well understood 
cosmological fluids in a natural way \cite{Hu:1998kj}.  

Using this phenomenological three parameter fluid description we investigated 
the effect of an imperfect dark energy fluid and of unified dark matter and
dark energy models on the matter power spectrum and on the CMBR 
temperature anisotropies. For most models we find that free streaming 
effects tend to smooth density fluctuations. However, there are some exceptions, 
described below.

In dark energy models where $-1\le w<0$, we found that increasing 
the anisotropic stress results in a swifter decay of dark energy 
overdensities, which is seen in the CMBR spectrum 
as an amplification of the ISW effect. The opposite occurs in the case of 
phantom dark energy ($w<-1$), for which the anisotropic stress supports the 
growth of overdensities and thus reduces the ISW effect. However, the 
impact 
of anisotropic stress on the CMBR spectrum can be closely mimicked by 
varying the sound speed of dark energy. This makes it difficult to 
distinguish between these two fluid properties. 

In addition, we found that negative sound speeds are also consistent with observations,
if shear viscosity is included. The situation that the pressure perturbation
(evaluated in the comoving frame) is of the opposite sign than the density perturbation,
is formally unproblematical to define, but when $c_{vis}^2=0$ it will exhibit unlimited
growth of density fluctuations. However, when $c_{vis}^2>0$ this does not 
occur. For a suitable 
choice of parameters a low amplitude for the CMBR quadrupole is produced,
in accordance with observations.

In models unifying dark matter and dark energy extended
with shear, it is found that the anisotropic
stress can stabilize the effect of the adiabatic pressure perturbation, thus 
slightly improving the compatibility 
of these models with large scale structure observations.
It remains to be seen how
one can loosen the constraints by  
allowing for an anisotropic stress. Our main objective here was
to use these models as examples of dark energy with evolving
$w$, $c_s$ and $c_{vis}$. The conclusion taken is that, in contrast to 
the simplest fluid models with constant $w$, $c_s$ and $c_{vis}$, 
in specific scenarios the shear stress can have consequences 
distinguishable with present observational data.
 
In general, we found that anisotropic perturbations in dark energy
is an interesting possibility which is not excluded by the present
day observational data. Furthermore, we found that the CMBR 
large scale temperature fluctuations, due to the 
the ISW effect, are a promising tool to constrain the possible imperfectness of the 
dark energy component. Even when the anisotropic stress cannot be directly measured, it 
can still bias measurements of other parameters, for instance 
the dark energy speed of sound or its equation of state.


\acknowledgments
We thank H. Kurki-Suonio for useful discussions. 
TK is supported by Magnus Ehrnrooth Foundation. 
DFM acknowledge support from the Research Council of Norway 
through project number 159637/V30.


\bibliography{refs10}

\end{document}